\documentclass[utf8,a4paper]{article}
 
\usepackage{url}
\usepackage{color}
\usepackage{babel}
\usepackage{natbib}
\usepackage{amsmath}
\usepackage{graphicx}

\usepackage[affil-it]{authblk}

\begin{document}
	
\title{Delivery strategies to improve piglets exposure  to oral antibiotics} 

\author{Noslen Hernández}
\author{Béatrice B. Roques}
\author{Marlène Z. Lacroix}
\author{Didier Concordet%
	\thanks{Electronic address: \texttt{didier.concordet@inrae.fr}; Corresponding author}}
\affil{INTHERES, Université de Toulouse, INRAe, ENVT, France}

\date{}

\maketitle

\begin{abstract}

The widespread practice of delivering antibiotics through drinking water to livestock leads to considerable variability in exposure levels among animals, raising concerns regarding disease outbreaks and the emergence of antibiotic resistance. This variability is primarily driven by three pivotal factors: fluctuations in drug concentration within water pipes, variances in drinking behavior among animals, and differences in individual pharmacokinetic parameters. This article introduces an approach aimed at improving medication distribution by customizing it according to the drinking patterns of pigs, without escalating the medication dose. As examples, we demonstrate that incorporating the drinking behavior into the delivery of amoxicillin results in an increase in the percentage of piglets reaching an AUC/MIC ratio greater than 25h. Specifically, with \textit{Pasteurella multocida}, the percentage rises from 30$\%$ to at least 60$\%$, while with \textit{Actinobacillus pleuropneumoniae}, it increases from 20$\%$ to more than 70$\%$.

\end{abstract}

\section{Introduction}
The use of antibiotics in group-raised animals has emerge as a significant concern, particularly in the context of pig farming. Antibiotics are typically administered to pigs in various ways. One prevalent approach involves adding antibiotics to the water they consume, offering a consistent and convenient method to administer treatment to large groups of pigs. However, this method leads to huge variability in exposure levels because animals do not consistently experience the same dosage of the medication over time. Consequently, some animals will be underexposed, while others will be overexposed \citep{little2021water}.  

The balance between under- and over-exposure of pigs to antibiotics is a crucial issue in modern livestock farming. 
While overexposure contributes to the alarming rise of antibiotic-resistant commensal bacteria, a nuanced focus on underexposure is equally crucial. Underexposure not only leads to the outbreak of diseases among the animals, causing suffering and economic losses for farmers, but also plays a pivotal role in fostering the development of antibiotic-resistant bacteria, posing a significant risk to public health \citep{ferran2019can}. Addressing this challenge requires the development of optimal delivery strategies that promote responsible antibiotic use in pig farms.

To initiate progress in this direction, it is necessary to delve deeper into understanding the causes of this variation in exposures. A recent study on pig exposure to antibiotics administered through drinking water \citep{little2019water} identified three factors contributing to this variability:
\begin{itemize}
	\item Variation arising from the actual concentration of the drug in the drinking water \citep{vandael2020stability} have demonstrated that there is a significant variation in the drug concentration at the end of the pipe, often not reaching the recommended therapeutic concentration range.
	\item Variation arising from the animals’ diverse drinking behaviours, i.e., the variation in the dose of medication actually consumed by each individual.
	\item Variation arising from the different pharmacokinetic parameters between pigs, resulting in different absorption and/or elimination of the antibiotic (clearance).
\end{itemize}

In a previous study \citep{chassan2021matters}, we have shown that the effective concentration of the drug in the pipes supplying the water to the drinkers and the animals' drinking behaviour are the most significant sources of variability in animal exposure to antibiotics. Inter-individual variability in clearance and absorption rates is a minor factor contributing to exposure variability.

The same study demonstrated that when amoxicillin or doxycycline is distributed in drinking water for three consecutive days, it is possible to achieve sufficient exposures for common bacteria (e.g., {\it Pasteurella multocida} and {\it Actinobacillus pleuropneumoniae} ) only in a small percentage of animals. As a result, the currently recommended doses or concentrations in the water do not adequately cover a significant portion of the individuals.

Several solutions can be considered to address this underexposure in the group. The simplest solution to avoid a large number of animals being underexposed is to increase the concentration of antibiotics in the water. However, this approach increases overall antibiotic consumption, which in turn promotes the emergence of resistance. The administration of a loading dose at the start of treatment could allow effective therapeutic concentrations to be reached more rapidly in a larger number of animals and could partially reduce inter-individual variability \citep{ferran2020levers}. However, this loading dose is only available for antibiotics with long terminal half-lives and good bioavailabilities. Another solution is to modify the formulation to extend its half-life, but this is challenging for medications distributed through drinking water. An alternative solution, not extensively covered in the literature, would be to adapt the distribution of medication based on the specific drinking behaviour of the group of pigs being treated. 

Specifically, we claim that increasing the concentration of medication in the drinking water when piglets come to drink leads to a corresponding increase in the amount of medication they ingest. Besides, augmenting the quantity of medication ingested by the animals results in an elevation of their exposure to the medication. Consequently, if antibiotics are distributed based on the drinking behaviour of the animals, a substantial percentage of them experience heightened exposure to the medication, while the daily dose remains unchanged. In this article, we propose optimal distribution strategies based on animals' drinking behaviour that increase the exposure of the group of animals in a way detailed hereafter.

\section{Material and methods}

Each pen, containing approximately 17 piglets, was equipped with two connected feed dispensers and a water dispenser. Water was distributed ad libitum. The weight, drinking times and corresponding water consumption were recorded for 918 piglets monitored from 28 to 70 days old. Each piglet was equipped with an (Radio-Frequency Identification) RFID sensor attached to its earring which allowed its identification at each recording. All piglets were subjected to the same husbandry conditions: same temperature and ventilation. The light was turned on at 9h00 and off at 17h00. At the end of the experiment, individual water consumption profiles as well as weight variations were available for the 30 days of recording.

\subsection{Model for water consumption}

The data corresponding to the water consumption of each pig can be viewed as a time series \citep{Kendall1976}. Within these series, occasional extreme values indicate water intake levels that is not consistent with physiological needs. These extreme values include not only water ingested by the pig but also water wasted often linked to pigs playing with water. In this section, we introduce a model aimed at detecting and replacing these excessive water consumption.  

As a first step, we employed an additive model represented by the equation 
\begin{equation}\label{eq:water_consumption}
	Q_{ik} = T_i(t_{ik}) + P_i(t_{ik}) + \varepsilon_{ik}.
\end{equation}
Here, $Q_{ik}$ denotes the observed volume of water consumed by the $i$-th piglet at time $t_{ik}$. This model decomposes the time series $Q_{ik}$ into three fundamental components \citep{montgomery2015introduction}. Firstly, the trend term $T_i$ quantifies the individual increase in water volume consumed over time, offering insight into the long-term underlying pattern in the data. Secondly, the seasonal or periodic term $P_i$ captures the repeatability of the drinking pattern from one day to the next, spotlighting the periodic fluctuations in the data that occurs with a fixed frequency. Lastly, the residual component $\varepsilon_{ik}$ encapsulates variations not explained by the trend and periodicity terms, encompassing the unpredictable aspects of the time series.   

For the estimation of the trend term $T_i$ we use a non-parametric regression technique called Locally estimated scatterplot smoothing (LOESS) that fits a series of weighted least square regressions to subsets of data, allowing the regression curve to vary flexibly across different regions of the time series. To facilitate the estimation of the periodic component $P_i$, the hours of the days were discretized into one-hour intervals. This discretization allows expressing $P_i(t)$ in the following form:
\begin{equation}
	P_i(t) = \sum_{j=0}^{23}c_{ij}1_{\{j\leq t-24[t/24] < j+1\}}
\end{equation}
where for any number $x$, $[x]$ refers to the integer part of $x$ and for any condition $E$, $1_{\{E\}}$ denotes the indicator function that takes the value 1 if the condition $E$ is true and 0 otherwise. The coefficients $c_{ij}$ provide insights into the hourly variations or patterns in water consumption over the course of a day. They quantify the degree to which each hour contributes to the periodic fluctuations observed in the volume of water consumed. Therefore, the estimation of the function $P_i(t)$ reduces to the estimation of the coefficients $c_{ij}$. 

It is important to note that the subscript $i$ in the model (\ref{eq:water_consumption}) corresponds to individual piglets, indicating that the model is fitted separately for each piglet's data. Thus, any parameters dependent on the piglet number $i$ may vary from one piglet to another. All estimations were done using the software R. 

In the second step, the value of $Q_{ik}$ was recognized as an outlier if its corresponding residual, $\widehat{\varepsilon_{ik}}$, exceeded a certain threshold. Specifically, when $Q_{ik}$ was greater than $\hat{T}i(t{ik}) + \hat{P}i(t{ik}) + 2\sqrt{\hat{\rm{var}}(\varepsilon_{ik})}$, it was considered an extreme value and replaced with $\hat{T}i(t{ik}) + \hat{P}i(t{ik}) + 2\sqrt{\hat{\rm{var}}(\varepsilon_{ik})}$. Conversely, if the water consumption was not identified as an extreme value, it remained unchanged and was retained for use in the subsequent PK stage.

\subsection{Model for pharmacokinetics}

The quantities of antibiotics to be diluted daily in the water to achieve a dosage of 20 mg/kg for amoxicillin and 10 mg/kg for doxycycline are typically determined based on the average weight of the animals and the average volume of water consumed throughout the day. Population kinetic models of amoxicillin and doxycycline have already been studied in pigs \citep{rey2014use,del2006interindividual}. A two-compartment model provides the best description of the kinetics of these two drugs. The plasma concentration $Z_t$ in a pig drinking a single dose $D$ of medication at time $t=0$ can be summarized as follows:
\begin{equation}
	Z_t = D(f(t,\psi) + \sigma f(t,\psi) \varepsilon_t ),
\end{equation}
where $\psi$ is the vector of individual PK parameters for the piglet, $f$ is the expected plasma concentration, $\sigma$ is a coefficient of variation of $Z_t$ and $\varepsilon_t$ is a standard normal random variable independent of $\psi$.

By utilizing individual water consumption profiles along with previously published kinetic models, it becomes possible to calculate the expected concentration profile during the specified treatment period. More specifically, let's denote $t_{ik}$ as the $k$-th drinking moment of pig $i$, and $D_{ik}$ as the dose of medication drunk at that moment. The expected drug concentration at time $t$ is given by:
\begin{equation}
	C_i(t) = \sum_k \left(D_{ik}(f((t-t_{ik})_+, \psi_i) + \sigma f((t- t_{ik})_+,\psi_i) \varepsilon_t)\right),
\end{equation}
where $\psi_i$ is the vector containing the PK parameters of the $i$-th piglet and $(a)_+ = a$ if $a>0$ and $(a)_+=0$ otherwise. For each piglet, 300 PK profiles were simulated.

The $k$-th dose of drug ingested by the $i$-th pig, $D_{ik}$, obviously depends on the volume of water, $Q_{ik}$, it consumed at that moment. More precisely, $D_{ik}=Q_{ik}\times F(t_{ik})$. In this equation, $F(t_{ik})$ represents the concentration of drug diluted in the drinking water at the drinking moment $t_{ik}$ of the pig. The current practice is to choose a constant dilution of the antibiotic in the water throughout the day so that the average dose is 20 mg/kg for amoxicillin and 10 mg/kg for doxycycline. This means that function $F$ is constant during each day but it changes every day according to the average piglet’s weight.   

To increase the animals' exposure during the treatment period, we will vary this concentration throughout the day and during the treatment duration while keeping the overall dose unchanged. Let $F_0(t)$ be the concentration, constant over a day, ensuring doses of 20 mg/kg for amoxicillin and 10 mg/kg for doxycycline, and let $F(t) = F_0(t)\times G(t)$, where $G(t)$ is the percentage of $F_0(t)$ allocated at time $t$ in such a way that $\int_0^{24} G(t)dt=1$.  

When the function $G(t)$ is constant, it means that the medication concentration in the drinking water remains constant throughout the day. On the other hand, if the function $G(t)$ changes over time, it allows for changes in the medication concentration in the drinking water within the same day. As it is practically difficult to continuously vary the concentration over time, we assume that the concentration can only change at specific predetermined hours. Specifically, we explored two forms for $G$,
\begin{equation}
	G_a(t) = \sum_{i=0}^{23} a_i 1_{\{i \leq t'< i+1\}},
\end{equation}
where $\sum_{i=0}^{23} a_i = 1$ and $t'= t-24[t/24]$ represents the hour within the day; and,
\begin{equation}
	G_b(t) = \sum_{i=0}^{5\times 24-1} b_i 1_{\{i \leq t < i+1\}},
\end{equation}
with $\sum_{i=0}^{5\times 24-1} b_i = 1$.

The first function, $G_a$, allows for changes in concentrations during the day, but these changes remain the same for each day of treatment. The second function, $G_b$, permits varying concentration during the day differently for each treatment day.

The way of distributing the medication in drinking water can be summarized as the selection of a particular function $G$, referred to as the antibiotic distribution strategy. The distribution strategy is therefore defined by choosing values for $a_0,a_1,...a_{23}$ or for $b_0,b_1,...b_{5\times 24-1}$. The function $G_a$ (respectively $G_b$) can be interpreted as the fraction of the daily dose (respectively, the fraction of the total dose over 5 days) to be given during each time interval. For example, the values of the coefficients that correspond to the usual strategy are $a_0=a_1=...=a_{23}=1/24$ and $b_0=b_1=...=b_{5\times 24-1} = 1/(5\times 24)$. Hereafter, we will name daily-fractionated strategy the function $G_a$ and multi-day cumulative strategy the function $G_b$. 

\subsection{Criteria to find optimal strategies}

To determine an optimal strategy, it is necessary to define the criterion that each type of strategy optimizes. All the criteria introduced in this article will depend on the area under the curve (AUC) of the average concentrations obtained throughout the entire treatment duration.
Let $AUC_i^j(a_0,...,a_{23})$ (respectively $AUC_i^j(b_0,...,b_{5\times 24-1})$) be the average AUC over 5 days of treatment starting on day $j$, considering the drinking behaviour of the $i$-th individual for a strategy $(a_0,a_1,...a_{23})$ (respectively $(b_0,b_1,...b_{5\times24-1})$). Formally, these AUCs are defined as follows:
\begin{equation}
	5\times AUC_i^j(a_0,...,a_{23}) = \int_{24j}^{24(j+5)}\sum_k Q_{ik} F_0(t_{ik})G_a(t_{ik})(f((t-t_{ik})_+,\psi_i) + \sigma f((t- t_{ik} )_+,\psi_i ) \varepsilon_t )dt
\end{equation}
and
\begin{equation}
	5\times AUC_i^j(b_0,...,b_{5\times 24-1}) = \int_{24j}^{24(j+5)} \sum_k Q_{ik}F_0(t_{ik})G_b(t_{ik})(f((t - t_{ik})_+,\psi_i) + \sigma f((t-t_{ik})_+,\psi_i) \varepsilon_t)dt
\end{equation}
Four optimality criteria have been chosen.

\subsubsection{Criterion 1: Minimum AUC}

With this criterion, we aim to find the strategy that maximizes the smallest (among individuals) average AUC over the treatment duration. Therefore, the piglet with the least exposure during the treatment period determines the optimum. More specifically, we need to solve the following two optimization problems for each treatment start day $j$:
\begin{equation}
	\arg\sup_{a_0\geq 0,...,a_{23}\geq 0} \inf_i AUC_i^j(a_0,...,a_{23})
\end{equation}
and
\begin{equation}
	\arg\sup_{b_0\geq 0,...,b_{5\times 24-1}\geq 0} \inf_i AUC_i^j(b_0,...,b_{5\times 24-1})
\end{equation}

\subsubsection{Criterion 2: 5th percentile of AUCs}

The second criterion consist on the 5th percentile of the average AUCs over the treatment duration. Seeking the strategy that maximizes this criterion involves disregarding the 5$\%$ of the lowest exposures and finding the strategy that maximizes the 95$\%$ highest exposures. Intuitively, by using this criterion it is assumed that these lower exposures will always be present and difficult to control. The optimization problems to be solved can be broken down into two steps.

First, for a given sequence $a_0,...,a_{23}$, we define the 5th percentile of AUCs, denoted as $x(a_0,...,a_{23})$, such that:
\begin{equation}
	\frac{1}{n} \sum_{i=1}^n 1_{\{AUC_i^j(a_0,...,a_{23}) \geq x(a_0,...,a_{23})\}} = 0.95
\end{equation}

The optimal strategy $(a_0,...,a_{23})$ is the one that solves the optimization problem
\begin{equation}
	\arg \sup_{\substack{a_0\geq 0,...,a_{23} \geq 0 \\ \sum_k a_k=1}} x(a_0,...,a_{23}).
\end{equation}

The formulation of the optimization of this criterion for the multi-day cumulative strategy (i.e., the optimization of $(b_j)$) immediately follows from the previous two equations.

\subsubsection{Criterion 3: Inverse of the coefficient of variation of AUCs}

Recall that the coefficient of variation ($CV$) of a random variable $X$ is defined as $CV(X)=SD(X)/E(X)$, where $SD(X)$ represents the standard deviation of $X$ and $E(X)$ represents its expected value. 

Consequently, maximizing the inverse of the coefficient of variation of the average AUCs over 5 days, or minimizing the coefficient of variation, aims to find the strategy that increases the average AUC calculated across all piglets while minimizing the dispersion of these AUCs.

Formally, if we denote as $\overline{AUC^j(a_0,...,a_{23})} = 1/n\sum_{i=1}^n AUC_i^j(a_0,...,a_{23})$ the average across piglets of the average AUCs, the optimal strategy for this criterion is obtained by solving the following optimization problem:
\begin{equation}
	\arg\sup_{\substack{a_0\geq 0,...,a_{23}\geq 0 \\ \sum_k a_k=1}} \frac{\overline{AUC^j(a_0,...,a_{23})}}{\sqrt{\sum_{i=1}^n\left(AUC_i^j(a_0,...,a_{23})-\overline{AUC^j(a_0,...,a_{23})}\right)^2}}
\end{equation}

The optimization problem to be solved for the case of the multi-day cumulative strategy (i.e., the optimization of $(b_j)$) is analogous to the previous one. 

\subsubsection{Criterion 4: The area under the curve of the probability of target attainment (PTA) for a bacterium}

The sensitivity of a bacterium to an antibiotic is characterized by the minimum inhibitory concentration (MIC). Since not all bacteria have the same MIC, the sensitivity of the strain is described by a distribution of MICs. In this study, we considered two bacterial families: \textit{Pasteurella multocida} and \textit{Actinobacillus pleuropneumoniae}, whose MIC distributions for amoxicillin and doxycycline can be found on the EUCAST website \citep{EUCAST}.

A global measure to quantify the exposure of a group of animals to an antibiotic for a given bacterium is to calculate the percentage of “meeting” between an individual and a bacterium for which the AUC/MIC is large enough. For the two antibiotics of interest in this article, it is widely accepted \citep{ambrose2000use} that a maximum number of individuals/bacteria should have an AUC/MIC greater than 25h or 48h . These two bounds could respectively guarantee a bacteriostatic and bactericidal effect \citep{lees2015pharmacokinetic,andes200713}. The criteria commonly employed are thus $P(AUC/MIC \geq 25)$ and  $P(AUC/MIC \geq 48)$. One should admit that even if they can guide the choice of a dosing regimen, these bounds are empirically derived. Therefore, other close values could be used. Instead, we propose calculating a general index indicating whether or not the chosen regimen can cover a sufficient percentage of individuals. That is,
\begin{equation}
	AUC_{PTA}(a_0,...,a_{23}) = \int_{0}^{100} P\left(\frac{AUC(a_0,...,a_{23})}{MIC} \geq x \right)dx,
\end{equation}
whose value increases with the exposure of a large percentage of individuals. The choice of integrating the probability from 0 to 100h is subjective. The optimal strategy for this criterion is obtained by solving the following optimization problem:
\begin{equation}
	\arg\sup_{\substack{a_0\geq 0,...,a_{23}\geq 0 \\ \sum_k a_k=1}} AUC_{PTA}(a_0,...,a_{23}).
\end{equation}
The formulation for the optimization using the multi-day cumulative strategy is analogous.

\section{Results}

\begin{figure}[htb]
	\begin{center}
		\includegraphics[width=0.9\linewidth]{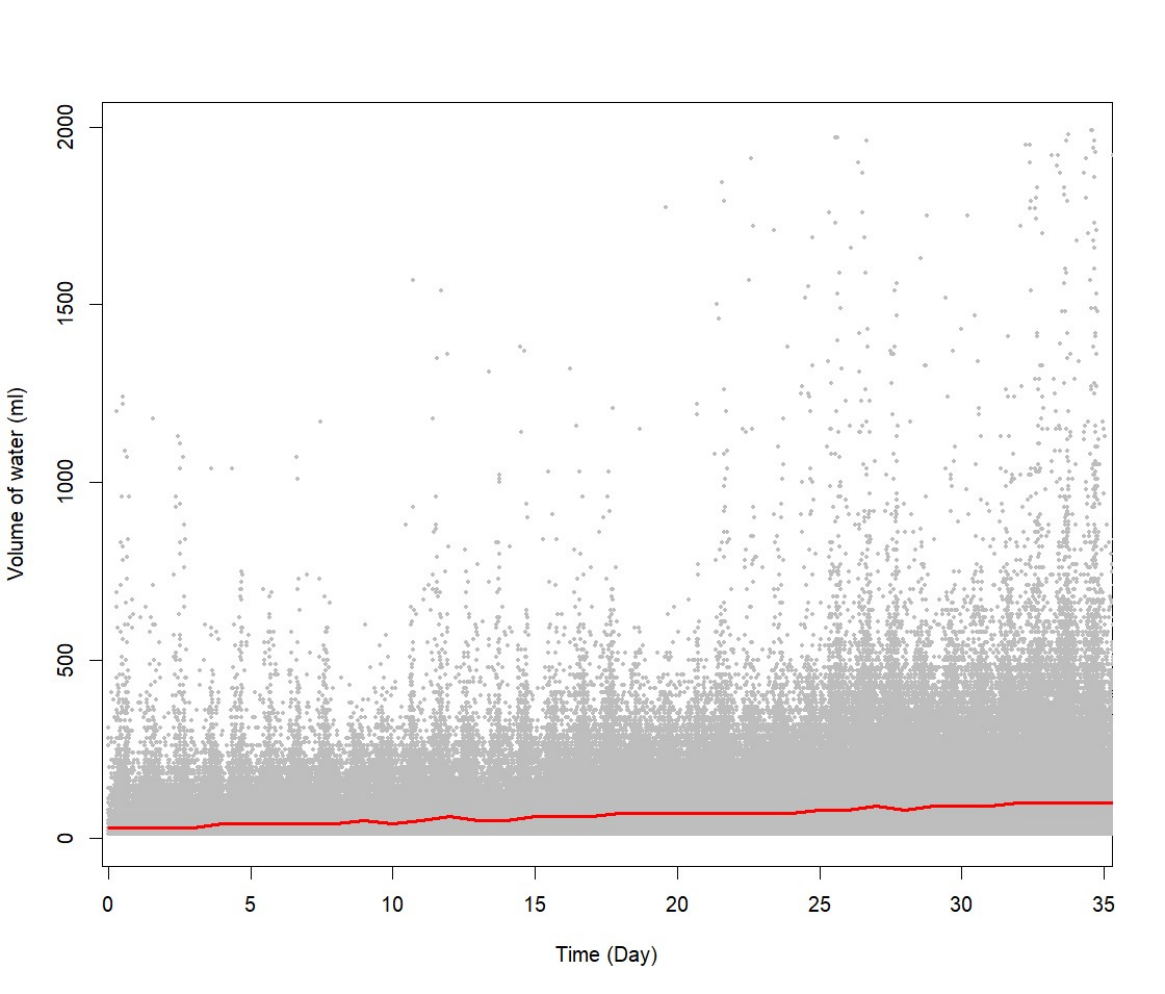}
	\end{center}
	\caption{Water consumption in millilitres of all piglets during the 35 days post-weaning. Each gray point indicates the volume of water consumed by a piglet on a specific day. The red line represents the median of the functions $T_i$ defined in Equation \ref{eq:water_consumption}.}\label{fig:water_consumption}
\end{figure}

Figure \ref{fig:water_consumption} shows the raw data concerning the water consumption of piglets during the 35 days post-weaning. Each gray point depicts the amount of water associated (either drunk or wasted) to a piglet on a specific day. Besides, the median of the functions $T_i$ obtained after fitting the model in Equation \ref{eq:water_consumption} to each piglet data is represented by a red line. The figure illustrates that the median water consumption increases over time (and with weight). In addition, a significant variability of the recorded volume of water is observed. However, the graph does not provide an assessment of the density of data points that refers to the percentage of piglets with low water intake and, consequently, resulting in a lower exposure. 

\begin{figure}[htb]
	\begin{center}
		\includegraphics[width=0.9\linewidth]{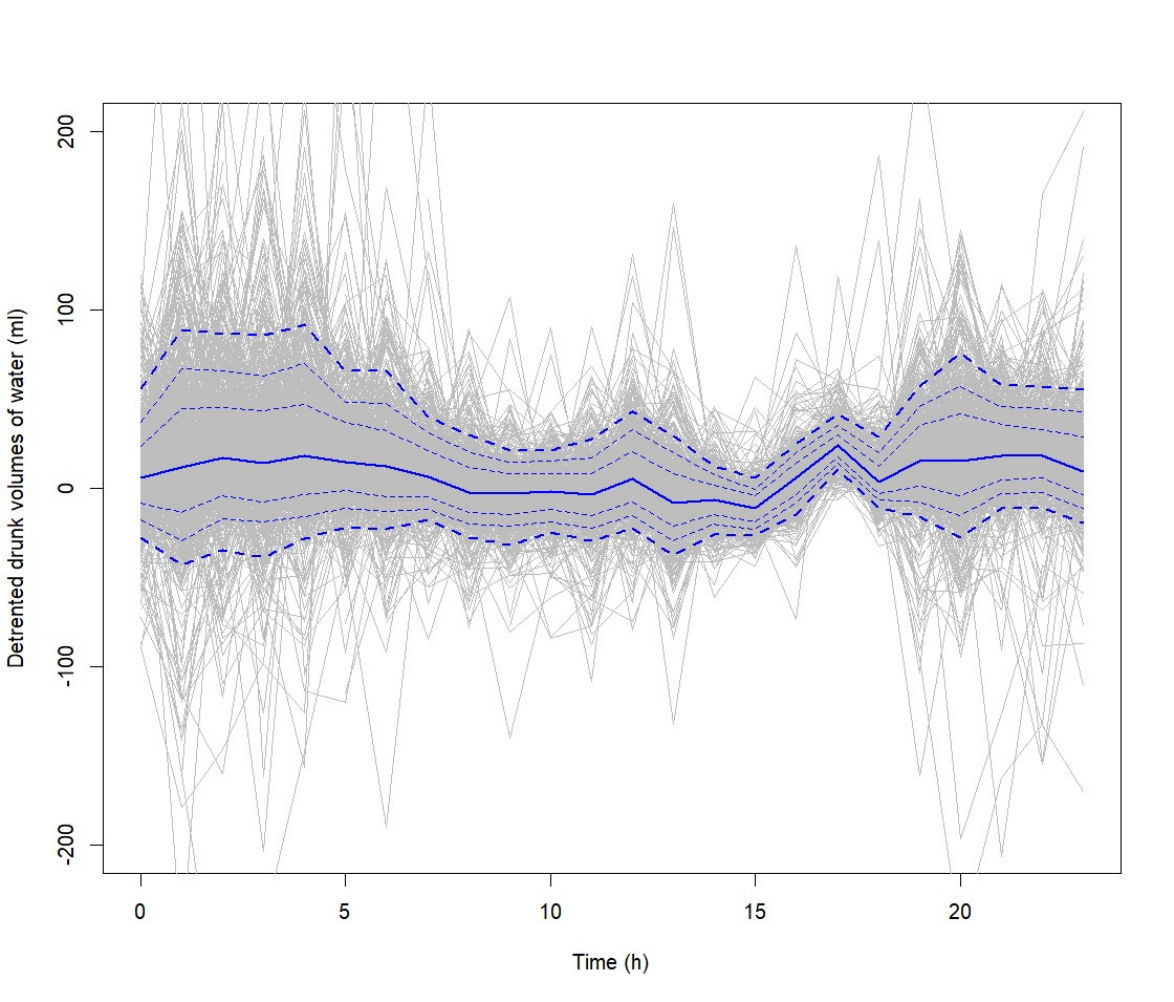}
	\end{center}
	\caption{For each individual i, the quantity $Q_{ik} - T_i(t_{ik}) = P_i(t_{ik}) + \varepsilon_{ik}$ as a function of the time of day is represented by a gray line. The blue curves represent respectively the 5\%, 10\%, 20\%, 50\%, 80\%, 90\% and 95\% quantile curves. }\label{fig:detrented_water_consumption}
\end{figure}

If for each piglet $i$, $T_i$ is removed from $Q_{ik}$, it is obtained $P_i(t_{ik}) + \varepsilon_{ik}$, which contains information about the 24 hours periodicity of the drinking behaviour. Figure \ref{fig:detrented_water_consumption} shows as gray lines the 24-hours chunks of $P_i(t_{ik}) + \varepsilon_{ik}$ for all the individuals as a function of the hours in a day. The high variability observed at a given time can be attributed to differences in the volume of water consumed by different animals at that moment and also to variations in the volume of water consumed by the same animal at that specific time across different days. On the contrary, a low variability at a given time indicates that, whatever the day, all the animals have the same drinking behaviour at that time. This plot reveals that from 7h to 19h, the dispersion of water quantities consumed by the animals is lower than that observed from 19h to 7h. Furthermore, during the period from 7h to 19h, it can be observed that the median, the 80\% and the 90\% exhibit local maxima showing that most animals consume more water around noon and around 17h.

\begin{figure}[htb]
	\begin{center}
		\includegraphics[width=0.9\linewidth]{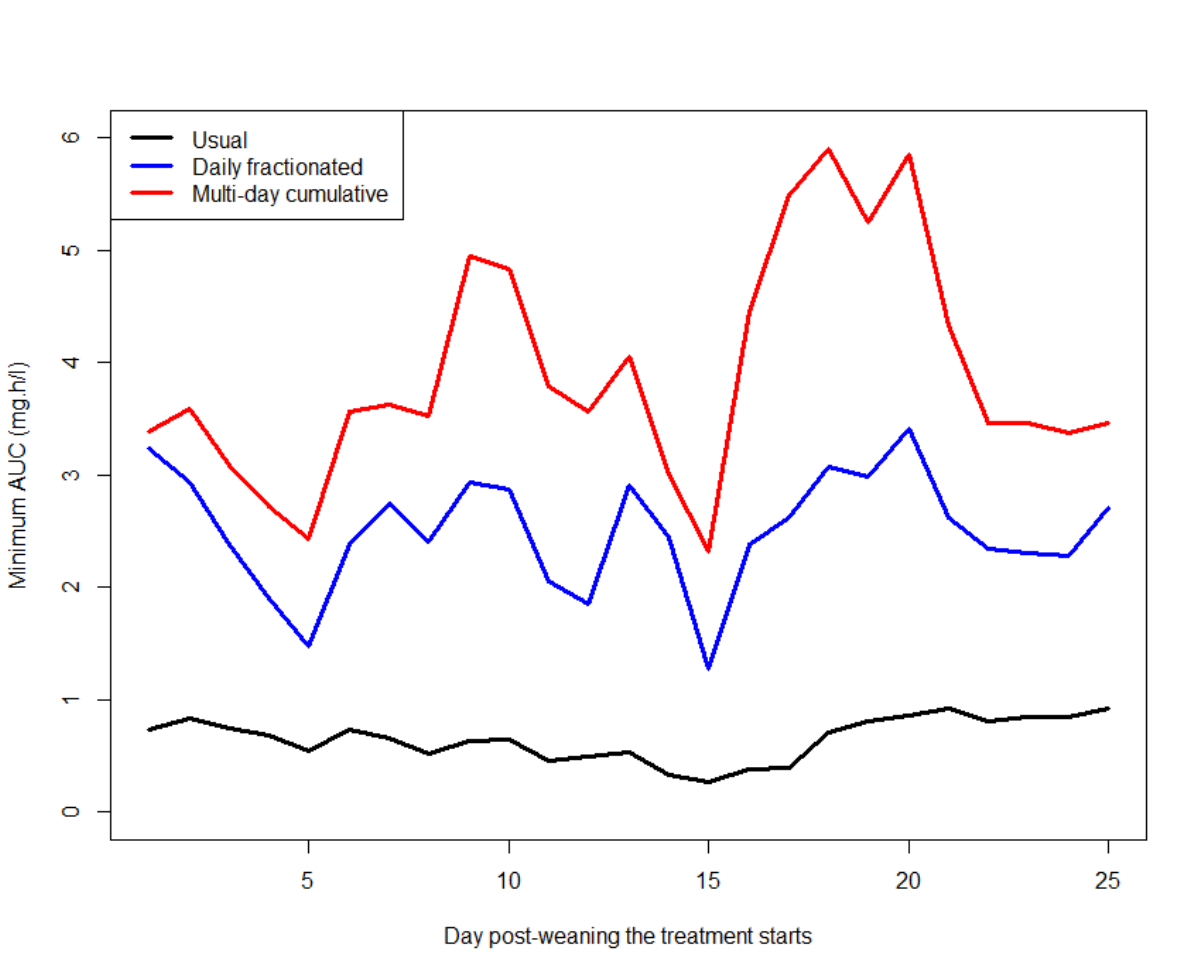}
	\end{center}
	\caption{Optimal value of criterion 1 (Minimum AUC) as a function of the days post-weaning the treatment starts for each of the three drug distribution strategies: usual (black line), daily-fractionated (blue line) and multi-day cumulative (red line). This computation corresponds to amoxicillin.}\label{fig:amox_minAUC}
\end{figure}

\begin{figure}[htb]
	\begin{center}
		\includegraphics[width=0.9\linewidth]{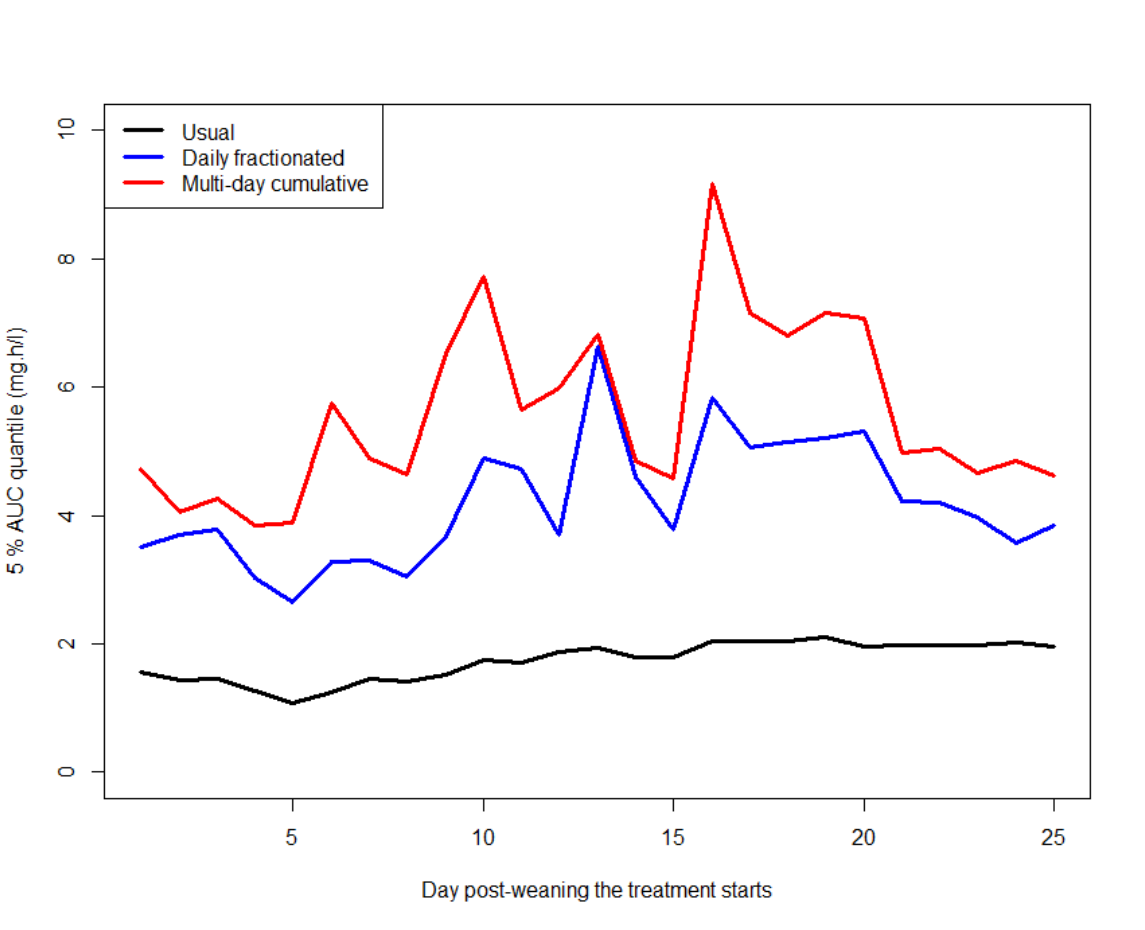}
	\end{center}
	\caption{Optimal value of criterion 2 (5\% percentile of AUC) as a function of the days post-weaning the treatment starts for each of the three drug distribution strategies: usual (black line), daily-fractionated (blue line) and multi-day cumulative (red line). This computation corresponds to amoxicillin.}\label{fig:amox_Q5}
\end{figure}

\begin{figure}[htb]
	\begin{center}
		\includegraphics[width=0.9\linewidth]{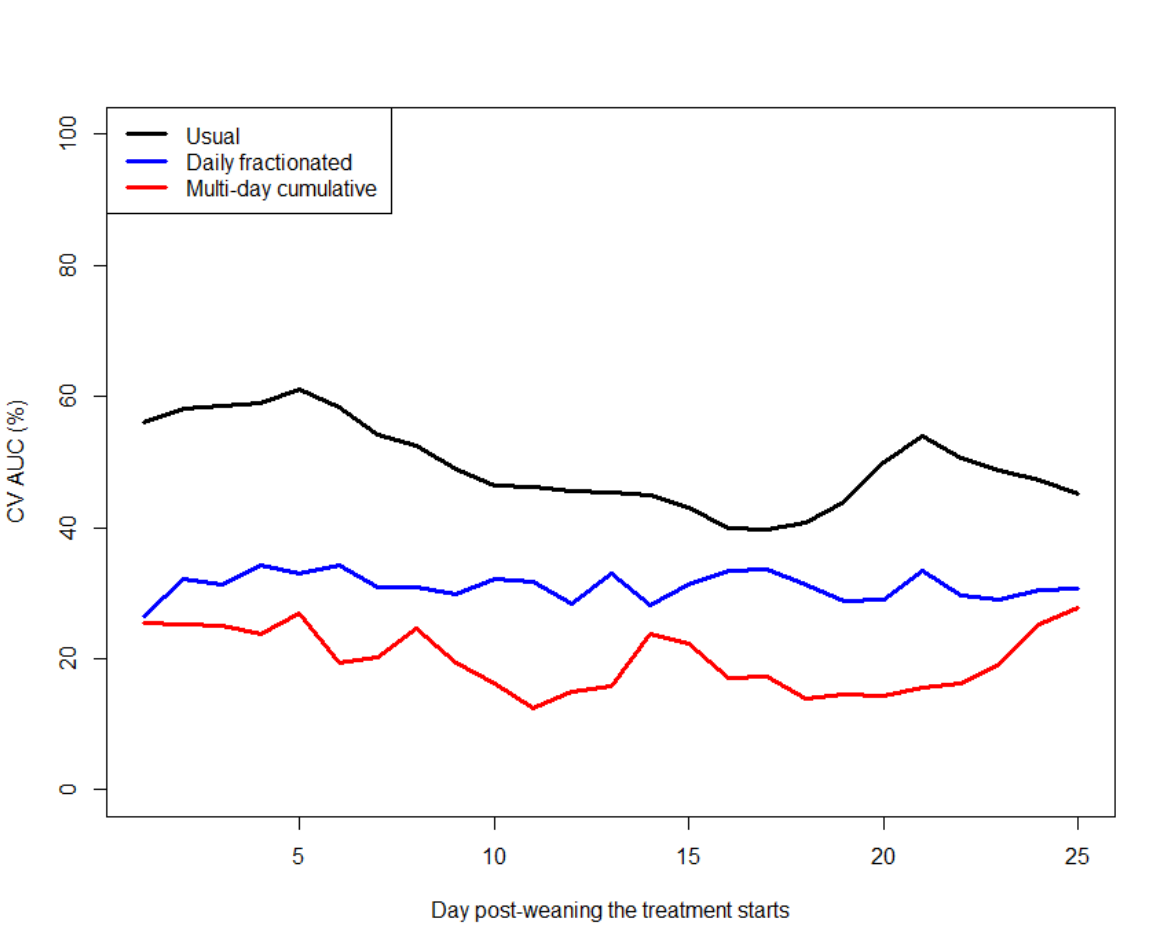}
	\end{center}
	\caption{Optimal value of criterion 3 (inverse of the coefficient of variation of AUC) as a function of the days post-weaning the treatment starts for each of the three drug distribution strategies: usual (black line), daily-fractionated (blue line) and multi-day cumulative (red line). This computation corresponds to amoxicillin.}\label{fig:amox_cv}
\end{figure}

For the results related to the computation of optimal strategies, we have opted to present in the primary text of this article only a subset of results concerning amoxicillin. The remaining findings concerning doxycycline are available as supplementary material (Figures S1-S9). Figures \ref{fig:amox_minAUC}, \ref{fig:amox_Q5} and \ref{fig:amox_cv}  illustrate the optimal value achieved for the criteria 1, 2 and 3, respectively, as a function of the day-post weaning for each type of distribution strategy (i.e., usual, daily-fractionated, multi-day cumulative). 

Figure \ref{fig:amox_minAUC} shows that using the usual strategy the minimum AUC is always less than 1mg.h/l whatever the day the treatment begins. Besides, the minimum AUC is always greater than 1 and can reach 3mg.h/l depending of the day the treatment is initiated when using the daily-fractionated strategy. Finally, the multi-day cumulative strategy allows to guarantee that, whatever the day the treatment starts, all piglets will have an AUC greater than 2.3-2.5 mg.h/l.

Figure \ref{fig:amox_Q5} shows that for the second criterion, using the usual strategy, 95\% of piglets will have an average AUC over 5 days approximately above 2 mg.h/l. This lower limit increases to 4 mg.h/l by using the multi-day cumulative strategy. 

Finally, for the third criterion, Figure \ref{fig:amox_cv} shows that, as expected, the multi-day cumulative strategy leads to lower CV than both the daily-fractionated and the usual strategies. Interpreting the value of CV in terms of individual exposure remains tricky.

\begin{figure}[htb]
	\begin{center}
		\includegraphics[width=0.9\linewidth]{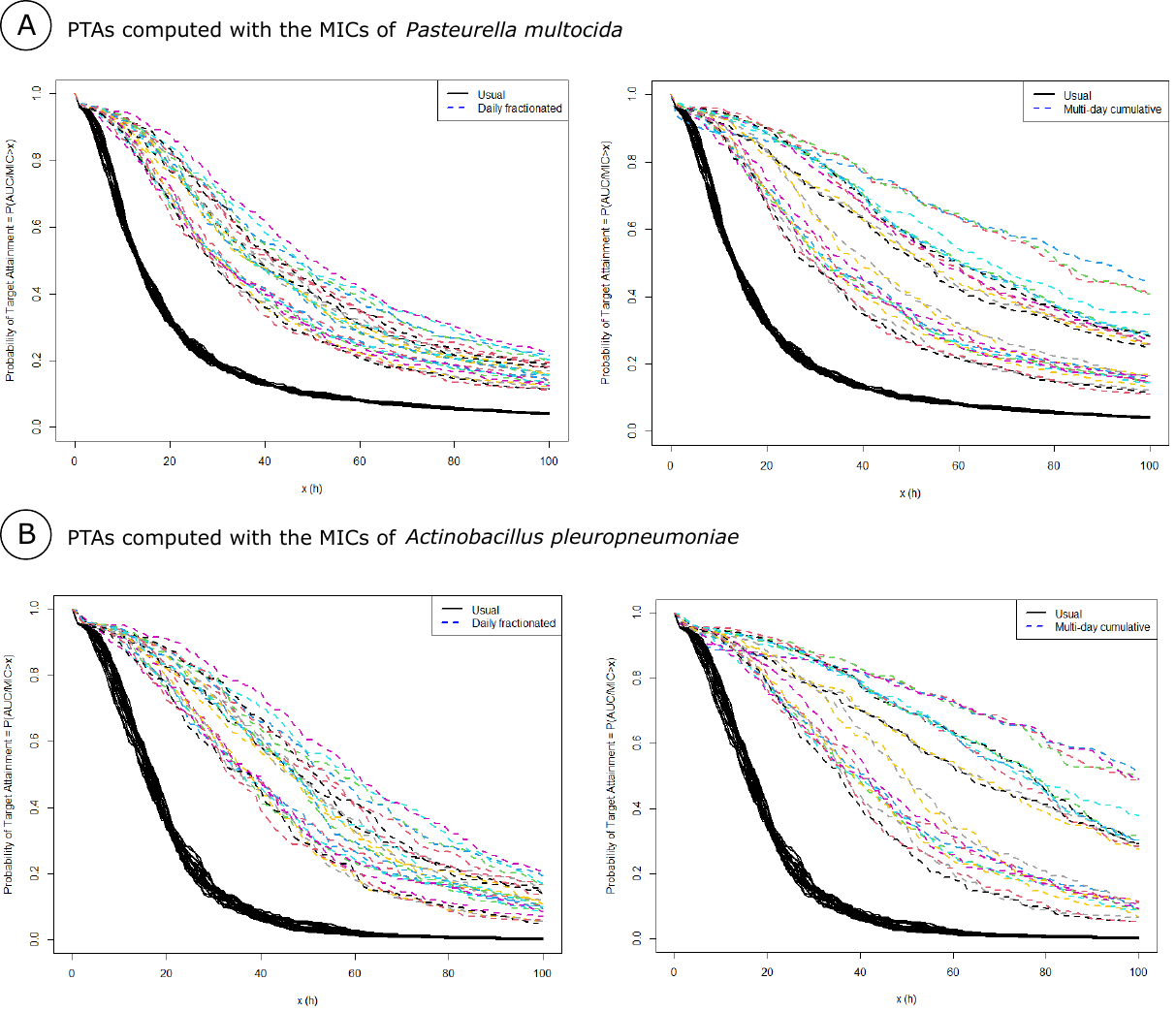}
	\end{center}
	\caption{(A) Probability of target attainment as a function of hours computed using the MICs of \textit{Pasteurella multocida}. Both panels show the curves of PTA obtained using the usual strategy (black plain lines). In addition, the left panel illustrates the curves of PTA obtained using the daily-fractionated strategy (dashed lines) and the right panel illustrates the curves of PTA obtained using the multi-day cumulative strategy (dashed lines). For each type of strategy, each curve correspond to a different starting day for the treatment. (B) Probability of target attainment as a function of hours computed using the MICs of \textit{Actinobacillus pleuropneumoniae}. Both panels show the curves of PTA obtained using the usual strategy (black plain lines). In addition, the left panel illustrates the curves of PTA obtained using the daily-fractionated strategy (dashed lines) and the right panel illustrates the curves of PTA obtained using the multi-day cumulative strategy (dashed lines). For each type of strategy, each curve correspond to a different starting day for the treatment.}\label{fig:amox_pta_AUC}
\end{figure}

Figure \ref{fig:amox_pta_AUC} shows the results concerning the optimization of criterion 4. This criterion involves the computation of the area under the  Probability of target attainment (PTA) curve. Figure \ref{fig:amox_pta_AUC}(A) shows the curves of PTA obtained with the MICs of\textit{ Pasteurella multocida }while Figure \ref{fig:amox_pta_AUC}(B) shows the curves obtained with the MICs of \textit{Actinobacillus pleuropneumoniae}.  

Both, Figure \ref{fig:amox_pta_AUC}(A) and Figure \ref{fig:amox_pta_AUC}(B) contains two panels. The left panel compares the curves of PTAs obtained using the usual strategy with the curves of PTAs obtained using the day-fractionated strategy. The right panel compares the curves of PTA obtained using the usual strategy with the curves of PTAs obtained using the multi-day cumulative strategy. The curves of PTAs obtained with the usual strategy are always depicted in black plain lines (one line corresponding to each day when the treatment begins). The curves of PTAs obtained using the day-fractionated and multi-day cumulative strategies are illustrated in dashed lines, each line representing also the day when the treatment starts.  

It can be observed that there is no dispersion between the curves of PTAs obtained using the usual delivery strategy. Whatever the day post-weaning the treatment begins, most of the piglets are equally under exposed. On the contrary, there is an important dispersion of the PTA curves depending on the day the treatment is initiated when the delivery of the drug is optimized. In all the cases, whatever the day the treatment begins, the PTA curve after optimization is above its counterpart without optimization.

\begin{figure}[h!]
	\begin{center}
		\includegraphics[width=\linewidth]{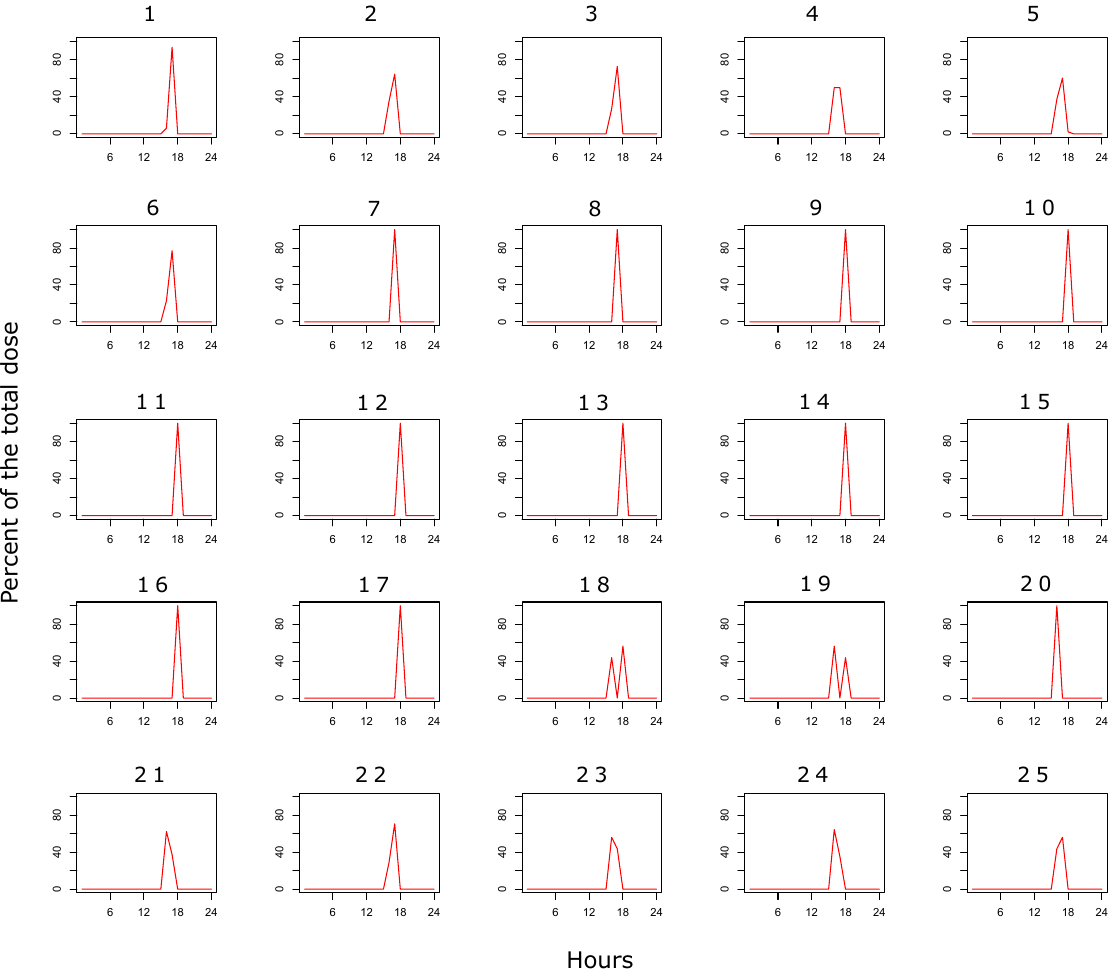}
	\end{center}
	\caption{Percentage of daily dose of amoxicillin to be delivered as a function of daytime for each day the treatment starts according to the daily-fractionated strategy when the criterion 4 is optimized using the MICs of \textit{Pasteurella multocida}.}\label{fig:amox_daily_optimal_time_PTA_pasteur}
\end{figure}

The result of the optimization problem for the case in which the daily-fractionated strategy and the PTAs computed with the MICs of \textit{Pasteurella multocida} are used is shown in Figure \ref{fig:amox_daily_optimal_time_PTA_pasteur}. This figure provides the distribution percentages of the daily dose to be dissolved in drinking water as a function of the time of day and the number of days post-weaning when the treatment commences. For instance, if the treatment starts 6 days post-weaning, 23\% of the daily dose should be mixed into the water provided between 15h and 16h, while the remaining 77\% of the dose should be dissolved in the water given between 16h and 17h. No medication should be administered during times not shown in the figure.

This figure shows that irrespective of the treatment initiation day, the optimal time range for delivering amoxicillin to achieve maximum AUC of the PTA with \textit{Pasteurella multocida} is between 15h and 18h.
 
\begin{figure}[htb]
	\begin{center}
		\includegraphics[width=\linewidth]{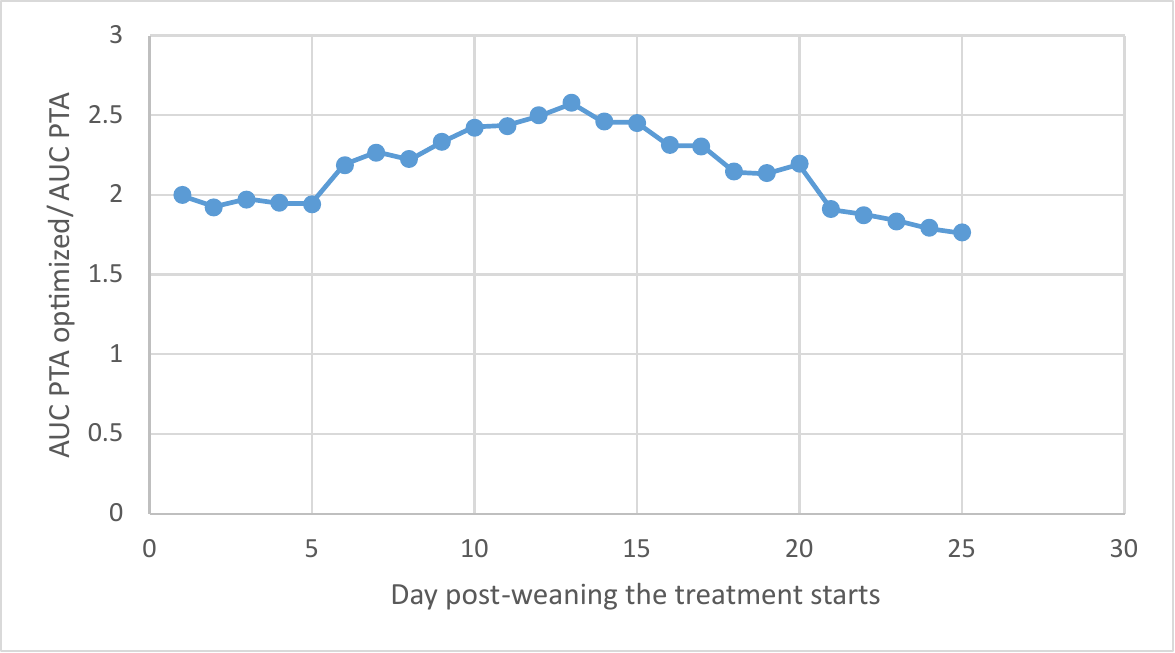}
	\end{center}
	\caption{Ratio of the are under the curve of PTA obtained using the daily-fractionated strategy and the MICs of \textit{Pasteurella multocida} as a function of the day the treatment starts.}\label{fig:ratio}
\end{figure}

Figure \ref{fig:ratio} shows the ratio between the area under the curve of PTA after using the daily-fractionated strategy and the corresponding area under the curve obtained with the usual strategy as a function of the day the treatment starts. A ratio equal to 2 means that the area under the curve of PTA after the optimization is twice the one without any optimization.

This graph indicates that the optimization increases the area under the curve of PTA by about a factor 2 (from 1.7 to 2.6). But the increase varies with the day the treatment is initiated. When the treatment is initiated less than 8 days after the post-weaning or at least 20 days after the post-weaning, the improvement of the area under the curve of PTA is about 2 and it can reach more than 2.5, 13 days after the post-weaning.

\begin{figure}[htb]
	\begin{center}
		\includegraphics[width=\linewidth]{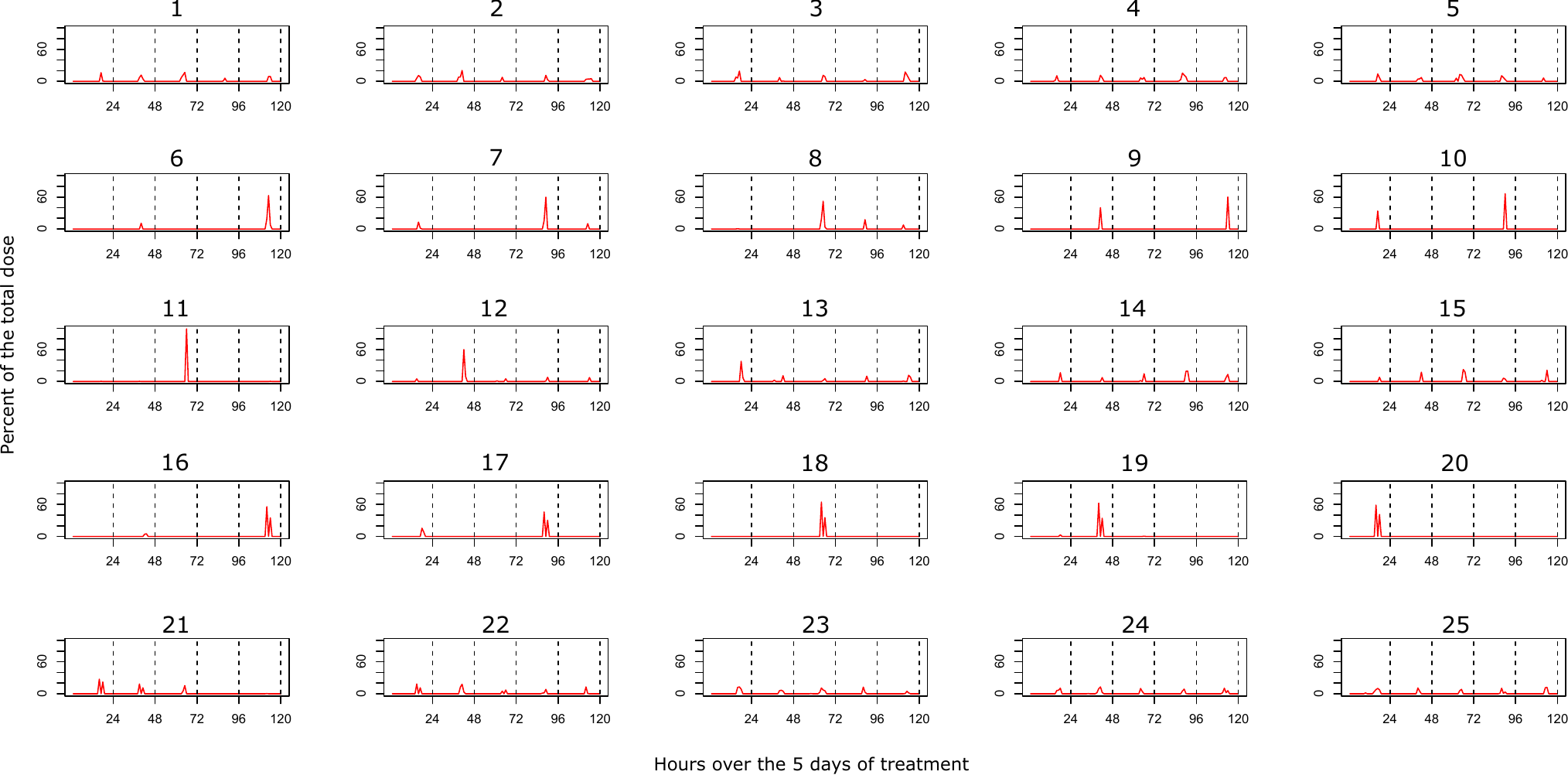}
	\end{center}
	\caption{Percentage of dose of amoxicillin to be delivered as a function of hours according to the multi-day cumulative strategy when the criterion 4 is optimized using the MICs of \textit{Pasteurella multocida}. Each plot correspond to a different day post-weaning the treatment starts. The vertical lines represent the end of each day of treatment.}\label{fig:amox_multiday_optimal_time_PTA_pasteurella}
\end{figure}

The result of the optimization problem when the multi-day cumulative strategy is used and the PTAs are computed using the MICs of \textit{Pasteurella multocida }is shown in Figure \ref{fig:amox_multiday_optimal_time_PTA_pasteurella}. Each plot in the figure corresponds to a starting day of treatment and shows the dose distributed during the 120 hours (5 days) of treatment. Remember that for this strategy, differently to the daily-fractionated strategy, it is obtained as a result of the optimization problem the optimal percent of total dose to be distributed across the 5 days of treatment.

\section{Discussion}

This article shows that the general exposure of a group of animals can be improved by adapting the way the drug is delivered to their dipsic behaviour. 

Husbandry conditions significantly influence pigs' drinking behavior. Factors like temperature, light exposure duration, and intensity can impact their circadian rhythm. Moreover, food type, quality, feeding schedules, pen space, and interactions with litter mates and other piglets all contribute to stress and potentially aggressive behavior. External elements such as ventilation, air quality, and sudden noises can also induce stress and anxiety in piglets \citep{pijpers1991influence, seddon2011can, turner1999influence,mroz1995water,vermeer2009motivation}. Due to these diverse influences, the findings in this article heavily rely on specific husbandry conditions and may not directly apply to other breeds. To implement the proposed strategy effectively, it's crucial to "calibrate" the approach according to the unique husbandry conditions of each breed.

However, even with these initial considerations, it becomes evident that the conventional antibiotic distribution strategy (called in this article usual strategy) appears to be, at first glance, among the least effective, despite its straightforward implementation. Labeling it as ``the least effective" means that any modification to this strategy enhances the animals' exposure to the medication.

We proposed a model for pig water consumption. Our primary aim in introducing this model was to address the issue of high water consumption values that do not align with the actual amount of water ingested by the animals. This empirical model aims at avoiding the use of the general rule of thumb that says that every day, a piglet drinks roughly 10\% of its body weight and never more than 20\% of its body weight. The introduced model is an empirical model aiming to describe “roughly” the animal drinking behaviour. We adopted a classical decomposition framework within time series analysis, which entails methods for breaking down a time series into its core components: trend, seasonal variation, and residuals. More sophisticate approaches to model water consumption have been introduced in the literature \cite{MADSEN2005138,bus2023circadian}. However, it’s worth noting that the choice of this model is entirely secondary in this paper and has minimal, if any, impact on the results we derived. Indeed, the criteria we examine (except for criterion 3) are constructed based on the lowest quantiles of the water consumption distribution. In essence, the criteria rely on piglets with lower water intake, consequently experiencing reduced exposure to medication. These criteria are extremely robust to outlier maximum values.

We have proposed the optimization of four criteria in order to obtain optimal antibiotic distribution strategies (i.e., minimum AUC, 5\% quantile of AUC, inverse of CV, AUC of PTA for a given bacterium). The introduced list of criteria is not exhaustive and many other criteria can be proposed. 

Once the responsible bacteria causing the infection are identified, using the Area Under the Curve (AUC) of Probability of Target Attainment (PTA) for a specific bacterium becomes the most suitable criterion. This is because PTA enables the computation of the percentage of individuals with an AUC/MIC exceeding the pharmacokinetic-pharmacodynamic cut-off \citep{toutain2019vetcast}. On the contrary, when the bacteria aren't identified, the most robust criterion, meaning the results are not affected by unreasonably high water consumption, is the 5\% quantile of AUC. This criterion is simple to implement. Another possibility could involve defining a criterion related to probabilistic antibiotherapy, which depends on the knowledge of the prevalence of various bacterial families in the specific breeding type under consideration.

In addition to the introduced criteria, we have proposed two types of strategy to model how the antibiotic is delivered throughout the treatment, namely the daily-fractionated strategy and the multi-day cumulative strategy. Our results indicate that the multi-day cumulative strategy generally provides greater exposure of the animals.       

Our results suggest that, regardless of the chosen criterion and the strategy, the concentration of the antibiotic in drinking water should be increased when the entire (or at least the vast majority) of animals drink water. It's widely acknowledged among pig farmers that pigs consume water around noon. Therefore, an intuitive idea could be to distribute the antibiotics at this time. Nevertheless, the obtained results indicate that this is not opportune in all the cases, suggesting other times for the distribution of the antibiotic. This can be explained by the fact that although several animals drink water around noon, the range of time in which they individually consume water around that time is wide. Therefore, there is no collective synchronization around that schedule. 

When using the daily-fractionated strategy, i.e., where identical antibiotic distribution pattern is maintained throughout the treatment days, it appears advantageous to administer the antibiotic between 15h and 18h (for both amoxicillin and doxycycline). This is quite consistent across all the criteria (Figures S10-S12 in the supplementary material show the optimal distribution pattern obtained using criterion 1, 2 and 3, respectively). Moreover, the selected optimal times do not vary significantly regardless of the day post-weaning the treatment starts.    

When using the multi-day cumulative strategy, i.e., one antibiotic distribution pattern is chosen across the 120 hours of treatment, we can see that the pattern varies depending on the day when the treatment starts. For some of the starting days of treatment (1, 2, 3, 4, 5, 14, 15, 22, 23, 24, 25) the optimal multi-day cumulative strategy is very close to the daily-fractionated strategy, with approximately the same pattern of antibiotic distribution for all treatment days. However, the result corresponding to other starting days of the treatment (6, 7, 8, 9, 10, 11, 12, 13, 16, 17, 18, 19, 20) suggests that the optimal strategy consists in administrating all the dose in one or two day of the five days of treatment. In fact, in all these cases, the days when the optimal strategy suggests administering all the medication are always day 10, 13 and 20. This is an indication that this can be a consequence of alterations (even imperceptible ones) in the management practices of the farm, which have influenced the dipsic behaviour of the piglets on these three days. We lack a reasonable explanation for this change of regime and we consider that this phenomenon deserves deeper investigation to be able to achieve a consistently optimal multi-day cumulative strategy regardless of the starting day of the treatment. However, putting the multi-day cumulative strategy into practice might pose complexity in its implementation. This approach implicitly assumes that the quantity of drug that reaches the drinking trough is close to the expected one meaning a good solubility and stability of the antibiotic and a low adsorption to the pipe \citep{georgaki2023qualitative,vandael2020stability}.

Considering the considerable anticipated benefits, potentially doubling the optimized criteria, it appears more advisable, initially, to explore the fundamental factors influencing dipsic behaviour.

All our calculations are based on the empirical cumulative distribution functions of the AUC at each time. With the number of sample that we have (approximately 1000 piglets) this approximation is very accurately and the statistics presented are subject to little sampling variability. Therefore, the sometimes erratic nature of the curves presented in the Figures \ref{fig:amox_minAUC}-\ref{fig:amox_cv} cannot be attributed to sampling variability. They are likely attributable to deterministic changes in drinking behaviour (which cause a change in the distribution).

In conclusion, our findings highlight that the conventional approach of administering antibiotics through drinking water is far from being optimal. This article proposes a methodology enabling each farm to increase the animal exposure without increasing antibiotic usage, based on a preliminary study characterizing the animals' drinking behavior.

All these outcomes stem from conducting group treatments, which involved maintaining identical concentrations of antibiotic in water for all animals. Individualizing the delivery of the drug could increase the individual exposure. This involves modelling the unique drinking behaviour of each animal and subsequently adjusting the antibiotic concentration in the water consumed by each individual. This approach parallels Therapeutic Drug Monitoring, used in human medicine for drugs with a low therapeutic index, as it closely aligns with constructing individual parameter estimates derived from population pharmacokinetic models.

\section*{Conflict of Interest Statement}
Noslen Hernández, Béatrice B. Roques, Marlène Z. Lacroix and Didier Concordet declare that they have no conflict of interest that might be relevant to the contents of this manuscript.

The authors affirm that the research was conducted without any commercial or financial affiliations that could be perceived as potential conflicts of interest.

\section*{Author Contributions}
DC proposed the methodology. NH and DC performed the analyses. NH, DC, BR, and ML drafted the paper. All authors critically reviewed several drafts and approved the final manuscript.

\section*{Funding}

The author(s) declare that no financial support was received for the research, authorship, and/or publication of this article. 

\section*{Data Availability Statement}

The data used in this study can be obtained upon request from the corresponding author. This data are not publicly accessible due to a confidentiality agreement.

\bibliographystyle{abbrvnat} 
\bibliography{biblio}

\begin{thebibliography}{23}
\providecommand{\natexlab}[1]{#1}
\providecommand{\url}[1]{\texttt{#1}}
\expandafter\ifx\csname urlstyle\endcsname\relax
  \providecommand{\doi}[1]{doi: #1}\else
  \providecommand{\doi}{doi: \begingroup \urlstyle{rm}\Url}\fi

\bibitem[Ambrose and Grasela(2000)]{ambrose2000use}
P.~G. Ambrose and D.~M. Grasela.
\newblock The use of monte carlo simulation to examine pharmacodynamic variance
  of drugs: fluoroquinolone pharmacodynamics against streptococcus pneumoniae.
\newblock \emph{Diagnostic microbiology and infectious disease}, 38\penalty0
  (3):\penalty0 151--157, 2000.

\bibitem[Andes and Craig(2007)]{andes200713}
D.~Andes and W.~A. Craig.
\newblock 13 pharmacokinetics and pharmacodynamics of tetracyclines.
\newblock \emph{in Theory and Clinical Practice}, page 267, 2007.

\bibitem[Bus et~al.(2023)Bus, Boumans, Engel, Te~Beest, Webb, and
  Bokkers]{bus2023circadian}
J.~D. Bus, I.~J. Boumans, J.~Engel, D.~E. Te~Beest, L.~E. Webb, and E.~A.
  Bokkers.
\newblock Circadian rhythms and diurnal patterns in the feed intake behaviour
  of growing-finishing pigs.
\newblock \emph{Scientific Reports}, 13\penalty0 (1):\penalty0 16021, 2023.

\bibitem[Chassan et~al.(2021)Chassan, H{\'e}monic, and
  Concordet]{chassan2021matters}
M.~Chassan, A.~H{\'e}monic, and D.~Concordet.
\newblock What matters in piglets’ exposure to antibiotics administered
  through drinking water?
\newblock \emph{Antibiotics}, 10\penalty0 (9):\penalty0 1067, 2021.

\bibitem[Del~Castillo et~al.(2006)Del~Castillo, Laroute, Pommier,
  Z{\'e}mirline, Ke{\"\i}ta, Concordet, and Toutain]{del2006interindividual}
J.~Del~Castillo, V.~Laroute, P.~Pommier, C.~Z{\'e}mirline, A.~Ke{\"\i}ta,
  D.~Concordet, and P.-L. Toutain.
\newblock Interindividual variability in plasma concentrations after systemic
  exposure of swine to dietary doxycycline supplied with and without
  paracetamol: a population pharmacokinetic approach.
\newblock \emph{Journal of animal science}, 84\penalty0 (11):\penalty0
  3155--3166, 2006.

\bibitem[EUCAST(2023)]{EUCAST}
EUCAST.
\newblock Antimicrobial wild type distributions of microorganisms, 2023.
\newblock URL \url{https://mic.eucast.org/search/}.

\bibitem[Ferran and Roques(2019)]{ferran2019can}
A.~A. Ferran and B.~B. Roques.
\newblock Can oral group medication be improved to reduce antimicrobial use?
\newblock \emph{The Veterinary Record}, 185\penalty0 (13):\penalty0 402, 2019.

\bibitem[Ferran et~al.(2020)Ferran, Lacroix, Bousquet-M{\'e}lou, Duhil, and
  Roques]{ferran2020levers}
A.~A. Ferran, M.~Z. Lacroix, A.~Bousquet-M{\'e}lou, I.~Duhil, and B.~B. Roques.
\newblock Levers to improve antibiotic treatment of lambs via drinking water in
  sheep fattening houses: the example of the sulfadimethoxine/trimethoprim
  combination.
\newblock \emph{Antibiotics}, 9\penalty0 (9):\penalty0 561, 2020.

\bibitem[Georgaki et~al.(2023)Georgaki, Vandael, Cardoso~de Carvalho~Ferreira,
  Filippitzi, De~Backer, Devreese, Dewulf, and
  Croubels]{georgaki2023qualitative}
D.~Georgaki, F.~Vandael, H.~Cardoso~de Carvalho~Ferreira, M.~E. Filippitzi,
  P.~De~Backer, M.~Devreese, J.~Dewulf, and S.~Croubels.
\newblock Qualitative risk assessment of homogeneity, stability, and residual
  concentrations of antimicrobials in medicated feed and drinking water in pig
  rearing.
\newblock \emph{BMC Veterinary Research}, 19\penalty0 (1):\penalty0 1--12,
  2023.

\bibitem[Kendall(1976)]{Kendall1976}
M.~Kendall.
\newblock \emph{Time Series. 2nd Edition}.
\newblock Charles Griffin and Co Ltd., London and High Wycombe, 1976.

\bibitem[Lees et~al.(2015)Lees, Pelligand, Illambas, Potter, Lacroix, Rycroft,
  and Toutain]{lees2015pharmacokinetic}
P.~Lees, L.~Pelligand, J.~Illambas, T.~Potter, M.~Lacroix, A.~Rycroft, and
  P.-L. Toutain.
\newblock Pharmacokinetic/pharmacodynamic integration and modelling of
  amoxicillin for the calf pathogens mannheimia haemolytica and pasteurella
  multocida.
\newblock \emph{Journal of Veterinary Pharmacology and Therapeutics},
  38\penalty0 (5):\penalty0 457--470, 2015.

\bibitem[Little et~al.(2019)Little, Crabb, Woodward, Browning, and
  Billman-Jacobe]{little2019water}
S.~Little, H.~Crabb, A.~Woodward, G.~Browning, and H.~Billman-Jacobe.
\newblock Water medication of growing pigs: sources of between-animal
  variability in systemic exposure to antimicrobials.
\newblock \emph{Animal}, 13\penalty0 (12):\penalty0 3031--3040, 2019.

\bibitem[Little et~al.(2021)Little, Woodward, Browning, and
  Billman-Jacobe]{little2021water}
S.~Little, A.~Woodward, G.~Browning, and H.~Billman-Jacobe.
\newblock In-water antibiotic dosing practices on pig farms.
\newblock \emph{Antibiotics}, 10\penalty0 (2):\penalty0 169, 2021.

\bibitem[Madsen and Kristensen(2005)]{MADSEN2005138}
T.~N. Madsen and A.~R. Kristensen.
\newblock A model for monitoring the condition of young pigs by their drinking
  behaviour.
\newblock \emph{Computers and Electronics in Agriculture}, 48\penalty0
  (2):\penalty0 138--154, 2005.
\newblock ISSN 0168-1699.
\newblock \doi{https://doi.org/10.1016/j.compag.2005.02.014}.
\newblock URL
  \url{https://www.sciencedirect.com/science/article/pii/S0168169905000475}.

\bibitem[Montgomery et~al.(2015)Montgomery, Jennings, and
  Kulahci]{montgomery2015introduction}
D.~C. Montgomery, C.~L. Jennings, and M.~Kulahci.
\newblock \emph{Introduction to time series analysis and forecasting}.
\newblock John Wiley \& Sons, 2015.

\bibitem[Mroz et~al.(1995)Mroz, Jongbloed, Lenis, and Vreman]{mroz1995water}
Z.~Mroz, A.~W. Jongbloed, N.~P. Lenis, and K.~Vreman.
\newblock Water in pig nutrition: physiology, allowances and environmental
  implications.
\newblock \emph{Nutrition Research Reviews}, 8\penalty0 (1):\penalty0 137--164,
  1995.

\bibitem[Pijpers et~al.(1991)Pijpers, Schoevers, Van~Gogh, van Leengoed,
  Visser, van Miert, and Verheijden]{pijpers1991influence}
A.~Pijpers, E.~Schoevers, H.~Van~Gogh, L.~van Leengoed, I.~Visser, A.~van
  Miert, and J.~Verheijden.
\newblock The influence of disease on feed and water consumption and on
  pharmacokinetics of orally administered oxytetracycline in pigs.
\newblock \emph{Journal of Animal Science}, 69\penalty0 (7):\penalty0
  2947--2954, 1991.

\bibitem[Rey et~al.(2014)Rey, Laffont, Croubels, De~Backer, Zemirline,
  Bousquet, Guyonnet, Ferran, Bousquet-Melou, and Toutain]{rey2014use}
J.~F. Rey, C.~M. Laffont, S.~Croubels, P.~De~Backer, C.~Zemirline, E.~Bousquet,
  J.~Guyonnet, A.~A. Ferran, A.~Bousquet-Melou, and P.-L. Toutain.
\newblock Use of monte carlo simulation to determine pharmacodynamic cutoffs of
  amoxicillin to establish a breakpoint for antimicrobial susceptibility
  testing in pigs.
\newblock \emph{American Journal of Veterinary Research}, 75\penalty0
  (2):\penalty0 124--131, 2014.

\bibitem[Seddon et~al.(2011)Seddon, Farrow, Guy, and Edwards]{seddon2011can}
Y.~Seddon, M.~Farrow, J.~Guy, and S.~Edwards.
\newblock Can monitoring water consumption at pen level detect changes in
  health and welfare in small groups of pigs.
\newblock In \emph{Proceedings of the 5th International Conference on the
  Assessment of Animal Welfare at Farm and Group Level. Ontario, Canada}, pages
  8--11, 2011.

\bibitem[Toutain et~al.(2019)Toutain, Sidhu, Lees, Rassouli, and
  Pelligand]{toutain2019vetcast}
P.-L. Toutain, P.~K. Sidhu, P.~Lees, A.~Rassouli, and L.~Pelligand.
\newblock Vetcast method for determination of the
  pharmacokinetic-pharmacodynamic cut-off values of a long-acting formulation
  of florfenicol to support clinical breakpoints for florfenicol antimicrobial
  susceptibility testing in cattle.
\newblock \emph{Frontiers in Microbiology}, 10:\penalty0 1310, 2019.

\bibitem[Turner et~al.(1999)Turner, Edwards, and Bland]{turner1999influence}
S.~Turner, S.~Edwards, and V.~Bland.
\newblock The influence of drinker allocation and group size on the drinking
  behaviour, welfare and production of growing pigs.
\newblock \emph{Animal Science}, 68\penalty0 (4):\penalty0 617--624, 1999.

\bibitem[Vandael et~al.(2020)Vandael, de~Carvalho~Ferreira, Devreese, Dewulf,
  Daeseleire, Eeckhout, and Croubels]{vandael2020stability}
F.~Vandael, H.~C. de~Carvalho~Ferreira, M.~Devreese, J.~Dewulf, E.~Daeseleire,
  M.~Eeckhout, and S.~Croubels.
\newblock Stability, homogeneity and carry-over of amoxicillin, doxycycline,
  florfenicol and flubendazole in medicated feed and drinking water on 24 pig
  farms.
\newblock \emph{Antibiotics}, 9\penalty0 (9):\penalty0 563, 2020.

\bibitem[Vermeer et~al.(2009)Vermeer, Kuijken, and
  Spoolder]{vermeer2009motivation}
H.~M. Vermeer, N.~Kuijken, and H.~A. Spoolder.
\newblock Motivation for additional water use of growing-finishing pigs.
\newblock \emph{Livestock Science}, 124\penalty0 (1-3):\penalty0 112--118,
  2009.

\end{thebibliography}

\end{document}